\documentclass[preprint,12pt]{elsarticle}

\usepackage{amssymb}
\usepackage{amsmath}

\journal{Journal of Subatomic Particles and Cosmology}

\begin{document}

\begin{frontmatter}



\title{Thirty years after the discovery of the top quark: the field enters an age of refinement and subtlety}


\author{Wolfgang Wagner} 

\affiliation{organization={University of Wuppertal, School of Mathematics and Natural Sciences},
            addressline={Gaußstraße 20}, 
            city={Wuppertal},
            postcode={42119}, 
            state={Nordrhein-Westfalen},
            country={Germany}}

\begin{abstract}
Thirty years after the first observation of on-shell top quarks the investigation of 
the heaviest elementary particle remains a thriving field of basic research, as was 
illustrated by the 18\(^\mathrm{th}\) edition of the annual Workshop on Top-Quark 
Physics hosted by Hanyang University in Seoul, Korea. 
Observing new scattering processes involving top quarks, precision measurements of top-quark 
properties, and the usage of top quarks as a means of exploration remain key elements 
of research, but are most recently complemented by the observation of even more subtle 
effects based on the application of refined experimental techniques.  
Based on the selection made in the experimental summary talk, 
this article highlights the most striking experimental results presented at the conference.
 \end{abstract}

\begin{keyword}
top quark \sep Large Hadron Collider 
\sep cross-section measurements
\sep top-quark pole mass
\sep bound-state effects 
\sep entanglement \sep flavour-changing neutral currents 
\sep charged-lepton-flavour universality



\end{keyword}

\end{frontmatter}


The year 2025 marks the thirtieth anniversary of the first direct observation of 
on-shell top-quark production at the Fermilab Tevatron by the CDF~\cite{CDF:1995wbb} 
and DØ Collaborations~\cite{D0:1995jca}. Since that time the field of top-quark physics 
has thrived and matured. Many more scattering processes involving the production of on-shell 
top quarks were observed, examples are the observation of single-top-quark 
production~\cite{D0:2009isq,CDF:2009itk} and the observation of the production of a 
top-quark-antiquark (\(t\bar{t}\)) pair in association with a \(Z\) boson or a \(W\) 
boson~\cite{TOPQ-2013-05,CMS-TOP-14-021}. 
Detailed investigations of top-quark production were 
performed by measuring a wealth of differential distributions and several production 
asymmetries; the top-quark decay was explored by measuring the helicity fractions of 
the \(W\) bosons occuring in the decay; top quarks were used to search for new particles 
and for scattering processes beyond the standard model of particle physics (SM).
At the Large Hadron Collider (LHC) top-quark physics has entered the era of precision 
measurements, most notably illustrated by the precise determination of the top-quark mass, 
as realised for example by the measurement of the CMS Collaboration in 
2016~\cite{CMS-TOP-14-022}.

At the 18\(^\mathrm{th}\) edition of the annual Workshop on Top-Quark Physics, a large number of 
impressing recent experimental results on the top quark were presented. This article 
highlights the most striking measurements presented at the conference and orders them along 
appropriate themes. While the themes of 
"discovery" (see Section~\ref{sec:observations}), precision (Section~\ref{sec:precision}) and 
exploration (Section~\ref{sec:exploration}), remain important, the measurements at the LHC have reached 
a level of refinement and subtlety which was not anticipated a few years ago. This includes 
in particular the observation of entangled (\(t\bar{t}\)) pairs and the observation of 
quasi-bound-state effects at the kinematic threshold of (\(t\bar{t}\)) production 
(Section~\ref{sec:subtlety}). 

\section{New observations}
\label{sec:observations}
The ATLAS Collaboration established the first observation of the associated production of a 
\(t\bar{t}\) pair and two high-\(p_\mathrm{T}\) photons~\cite{TOPQ-2023-03}. 
A Boosted Decision Tree was trained with 19 input variables to separate signal and 
background events. A fiducial cross-section of 
\(\sigma(t\bar{t}\gamma\gamma)=2.42^{+0.58}_{-0.53}\,\)fb is measured and agrees 
with the SM prediction within uncertainties. The significance of the new signal is \(5.2\) 
standard deviations.

The CMS Collaboration reported the first obseravtion of a single top quark in association 
with a \(W\) boson and a \(Z\) boson (\(tWZ\) production)~\cite{CMS-TOP-24-009}. The analysis 
employs a classifier based on a particle transformer to separate signal from background events, 
and makes use a Run~2 and Run~3 data, amounting to an integrated luminosity of 
\(200\,\)fb\(^{-1}\). The observed significance of the signal is \(5.8\) standard deviations, 
while \(3.5\) standard deviations are expected.

\section{Precision measurements}
\label{sec:precision}
Precision measurements at a hadron collider are a very challenging endeavour and strongly depend 
on advanced calibration techniques. The ATLAS Collaboration employs a new approach for the in-situ 
calibration of the jet energy scale. The combination of the \(E/p\) ratio with traditional 
\(p_\mathrm{T}\)-balance methods leads to a reduction of the uncertainties across the full  
\(p_\mathrm{T}\) range of the jets~\cite{JETM-2022-06}. The uncertainty in the determination 
of the integrated luminosity of a particular data set is a crucial element of precise cross-section 
measurements. The final determination of the integrated luminosity of the Run~2 data set 
recorded with the CMS detector yields \(137.88\pm1.01\,\)fb\(^{-1}\), corresponding to a relative  
uncertainty of \(0.73\)\%~\cite{CMS:2025wux}. A key element to reach this impressing precision is the usage of multiple 
luminometers to track the time dependence of the luminosity calibration.

The most recent determination of the top-quark mass, \(m_t\), based on top-quark decay properties 
by the ATLAS Collaboration is performed with a sample enriched in \(t\bar{t}\) events exhibiting a 
boosted top-quark topology~\cite{TOPQ-2022-24}. The analysis targets the semileptonic decay channel 
and employs the average mass of top-quark-jet, \(\bar{m}_J\), originating from the hadronically 
decaying top quark as an observable. A linear relation between \(\bar{m}_J\) and \(m_t\) is 
exploited for the measurement, yielding a value of \(m_t=172.95\pm0.53\,\)GeV. Two jet-related 
auxiliary variables are used to constrain the two leading systematic uncertainties, namely the 
uncertainty in the jet energy scale and the uncertainty in the recoil scheme employed 
when modelling gluon emissions from \(b\)-quarks produced in the top-quark decay. This recent 
measurement constitutes the most precise determination of \(m_t\) by a single measurement of the 
ATLAS Collaboration. Currently, the most precise value of \(m_t\) is given by the legacy 
combination of various measurements by the ATLAS and CMS Collaborations in Run~1 of the LHC, 
yielding \(m_t=172.52\pm0.33\,\)GeV~\cite{TOPQ-2019-13}.  

The ATLAS Collaboration further exploits the production properties of \(t\bar{t}\) events with an 
additional high-\(p_\mathrm{T}\) jet to determine the pole mass of the top 
quark~\cite{TOPQ-2018-23}. The sensitivity mainly arises from the production of these events at the 
kinematic threshold. The obtained value is \(m_t=170.73^{+1.47}_{-1.44}\,\)GeV.

Another precision measurement presented by the ATLAS Collaboration concerns the determination of 
the total, the fiducial and various differential cross-sections of \(t\bar{t}\) production in 
the dileptonic decay channel~\cite{TOPQ-2024-12}. The measurements employ the legacy 
determination of the integrated luminosity, reaching a relative precision of 
\(\Delta\mathcal{L}/\mathcal{L}=\pm0.83\%\), and supersede previous results~\cite{TOPQ-2018-26}. 
The analysis uses a new baseline sample of generated \(t\bar{t}\) events which features a 
matrix-element calculation at next-to-next-to-leading order in quantum chromodynamics matched 
to a parton-shower generator, 
implemented in the \textsc{Powheg}~\cite{Nason:2004rx,Frixione:2007vw,Alioli:2010xd} event generator 
based on the MiNNLO\(_\mathrm{PS}\) procedure~\cite{Mazzitelli:2020jio}. 
The new model significantly improves the description of the 
\(p_\mathrm{T}\) spectrum of top quarks, and thus avoids the introduction of an additional 
uncertainty covering the mismodelling in previous versions of the analysis. Uncertainties related 
to the modelling of initial-state radiation and final-state radiation are also reduced due to the 
higher order of the matrix-element calculation. The total \(t\bar{t}\) cross-section is measured to be
\[\sigma(t\bar{t})=829.3\pm1.3\,(\mathrm{stat})\pm8.0\,(\mathrm{syst})\pm7.3\,(\mathrm{lumi})
  \pm1.9\,(\mathrm{beam})\;\mathrm{pb.}\]
The relative uncertainty in the measured value is 1.34\% (previously 1.80\%). 

\section{Measurements of exploration}
\label{sec:exploration}
The accurate reconstruction and identification of the detected objects is a crucial element for 
measurements of exploration. In recent years, the ATLAS and CMS Collaborations have made 
impressive progress in the flavour tagging of jets. New machine-learning techniques, such as 
graph neural networks and transformer-based models are instrumental in reducing misidentification 
rates at the same level of tagging 
efficiency~\cite{ATLAS:2019bwq,ATLAS:2021cxe,ATLAS:2023lwk,Qu:2022mxj}.
The ATLAS Collaboration uses a deep neural network to replace a simple likelihood-based method 
for electron identification, and thereby improves the background rejection by a factor of 
two~\cite{ATL-PHYS-PUB-2022-022}. The CMS Collaboration devises the new Pile-up Per Particle 
Identification (PUPPI) algorithm and brings down the jet-energy pile-up offset close to 
zero~\cite{CMS-DP-2024-039}. The PUPPI algorithm is the default for the analysis of Run~3 data. 

Top quarks enable measurements of exploration in three different directions: Measurements of SM 
processes are performed in extreme phase spaces to look for deviations from SM predictions. 
Top quarks occur in decays of new particles which are not part of the SM, such as leptoquarks, 
vector-like quarks, supersymmetric particles or heavy Higgs bosons. As such, top quarks form 
an important "tool" in direct searches for physics beyond the SM (BSM). And finally, top quarks 
enable indirect searches for BSM physics, for example for flavour-changing neutral currents (FCNC), 
lepton-flavour violation, CP violation or rare decays. 

\subsection{Measurements of Standard Model processes in extreme phase spaces}
The CMS Collaboration explores the phase space of \(t\bar{t}\) production in the dilepton final 
state by measuring the triple differential cross-section as a function of \(p_\mathrm{T}(t\bar{t})\),
\(m(t\bar{t})\) and \(y(t\bar{t})\)~\cite{CMS-TOP-20-006}. 
All MC-generator models show local trends with respect to the measured data.
A measurement of the differential cross-section in the variable
\[m_\mathrm{minimax}^{bl}\equiv\mathrm{min}
  \left\{\mathrm{max}\left(m^{b_1\ell_1},m^{b_2l_2}\right),
  \mathrm{max}\left(m^{b_1\ell_2},m^{b_2l_1}\right)\right\}
\]
by the ATLAS Collaboration in the dilepton channel explores off-shell contributions and 
the interference of amplitudes with two resonant top-quarks, 
a single resonant top-quark and without intermediate top quarks~\cite{TOPQ-2020-14}. 
The high \(m_\mathrm{minimax}^{bl}\) region, beyond approximately 200\(\,\)GeV, is especially important 
for BSM searches. The bb4l Monte-Carlo event generator~\cite{Jezo:2016ujg} is a dedicated tool to 
describe these effects accurately. To be used for an experimental analysis a dedicated prescription 
to estimate modelling uncertainties is needed. To address this issue the ATLAS Collaboration 
performed comprehensive studies~\cite{ATL-PHYS-PUB-2025-036}.

Both collaborations, ATLAS and CMS, established \(t\bar{t}\) production in lead-lead collisions at 
the LHC~\cite{HION-2022-10,CMS-PAS-HIN-24-021}. These efforts pursue the long term goal of using 
the jets from top-quark decay products as a tool to study the quark-gluon plasma. 

\begin{figure}[!t]
\centering
\includegraphics[width=0.65\textwidth]{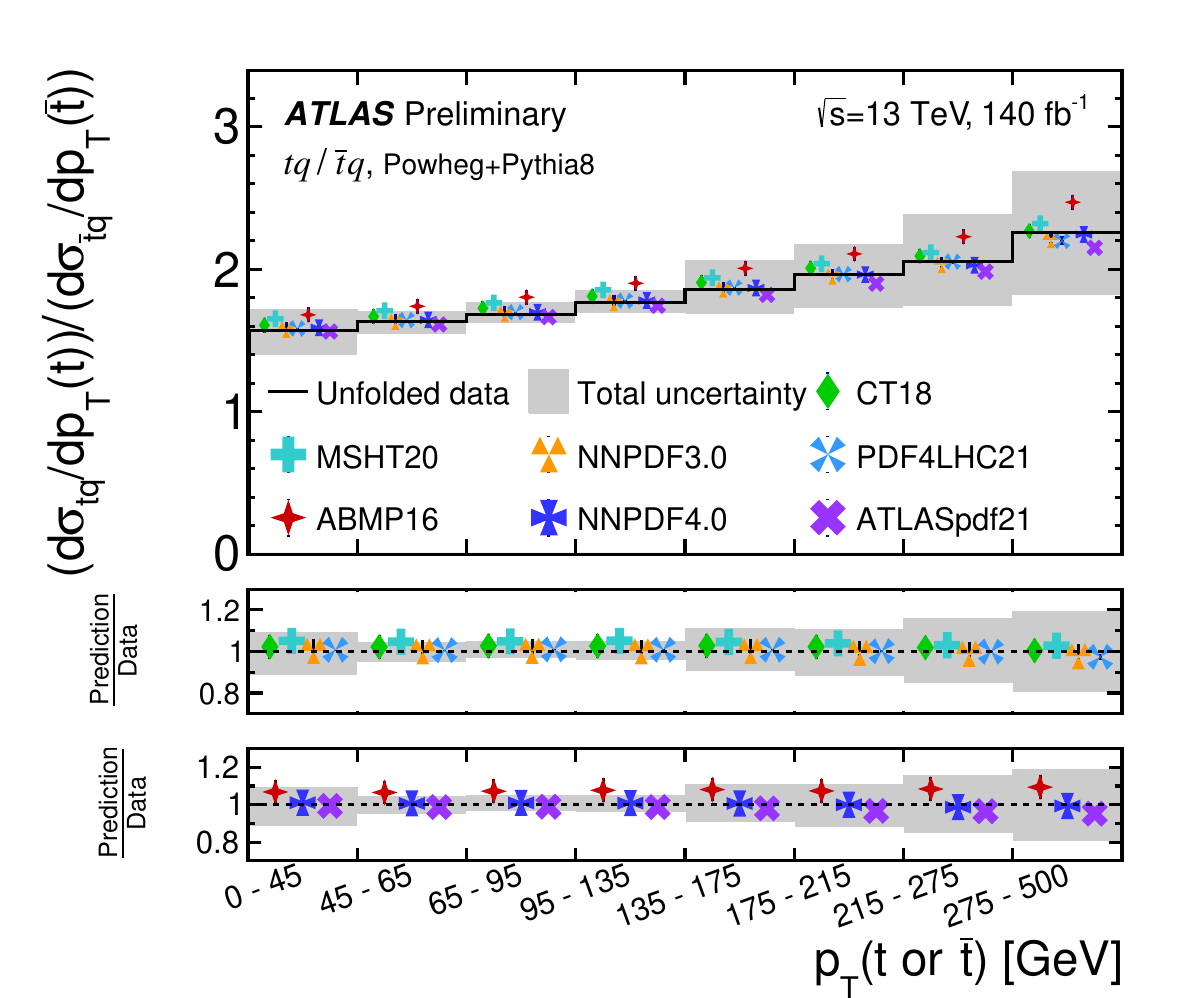}
\caption{Ratio of the unfolded differential cross-sections as a function of \(p_\mathrm{T}(t\;\mathrm{or}\; \bar{t})\). 
   The shaded band indicates the total measurement uncertainty. 
   The cross-sections are compared with theoretical predictions obtained from POWHEG+PYTHIA\(\,\)8 with different PDF sets. 
   The uncertainty on the theoretical predictions are the MC statistical uncertainties only.}
   \label{fig1}
\end{figure}

The ATLAS Collaboration measured differential cross-sections of single top-quark production in the 
\(t\)-channel (\(tq\) production) as a function of the top-quark \(p_\mathrm{T}\) and 
rapidity (\(y\))~\cite{ATLAS-CONF-2025-011}. Cross-sections of top-quark and top-antiquark 
production are separately measured. In addition, the ratio of the cross-sections is determined. 
The measurements are compared to the predictions by different event-generator setups, 
fixed-order calculations, and predictions based on different PDF sets. Figure~\ref{fig1} shows the 
measured ratio of differential cross-sections as a function of the \(p_\mathrm{T}\) of the 
top quark or top antiquark, respectively.
The cross-sections are compared with theoretical predictions obtained from POWHEG+PYTHIA 8 with different PDF sets.
The predictions obtained with the ABMP16 PDF set and, to a lesser extent 
the predictions obtained with the MSHT2020 and CT18 PDF sets overestimate the ratio of the diﬀerential
cross-section over the full measured range. The measurements of the differential cross-sections as a function of 
\(p_\mathrm{T}\) are further interpreted in an effective field theory (EFT) approach and limits on the coefficient of the 
four-quark operator \(O_{Q,q}^{3,1}\) are set.

The CMS Collaboration measured several differential cross-sections of \(t\bar{t}+W\) production in the 
two-lepton-same-charge-sign channel~\cite{CMS-TOP-24-003}. No significant trends are observed in any 
kinematic distributions. The ATLAS Collaboration explored the \(t\bar{t}+\ell^+\ell^-\) final state at high 
invariant mass, excluding the resonant region by requiring \(m(\ell^+\ell^-)>m_z+10\,\)GeV~\cite{TOPQ-2023-02}.
The observed signal strength in the signal region is well compatible with the expectation and limits 
are set on the coefficients of various EFT operators. Both, the ATLAS and CMS Collaborations, presented new 
measurements of the associated production of a Higgs boson and a \(t\bar{t}\) pair where the Higgs boson decays 
to a \(b\bar{b}\) pair~\cite{CMS-HIG-24-018,HIGG-2020-24}. Significant improvements in modelling the notoriously 
difficult \(t\bar{t}\)+jets background were implemented by constraining the flavour composition of the additional 
jets in dedicated control regions. 

In the future, advanced unfolding techniques have the potential to further boost the sensitivity of differential 
cross-section measurements. One opportunity is unbinned unfolding as implemented in the Omnifold 
package~\cite{Andreassen:2019cjw} which was used by the ATLAS Collaboration to characterise kinematic distributions 
of \(Z+\)jets production~\cite{STDM-2020-17}. Another possibility is simulation-based inference~\cite{Brehmer:2020cvb} 
aiming for the determination of true likelihood ratios.

\subsection{Direct searches}
Searches were performed for vector-like quarks which decay to a \(Wb\) final state~\cite{EXOT-2018-60} or 
to a top quark and a scalar particle decaying to a \(b\bar{b}\) pair~\cite{CMS-B2G-23-009}, 
where the scalar is either an SM Higgs boson or a BSM particle. 
No excess above the SM expectation is found and lower limits are set on the masses of the vector-like quarks 
up to 2.6\(\,\)TeV.

\subsection{Indirect searches}
The ATLAS Collaboration measured the quadrupule differential decay rate of singly produced top quarks. 
In this approach, the decay rate is expressed as an M-function expansion and the corresponding coefficients are 
determined. Based on these measurements limits on BSM contributions to the top-quark decay via four different EFT 
operators are set~\cite{TOPQ-2020-17}. Lepton-flavour universality is tested in \(W\) decays utilizing on-shell 
\(W\) bosons originating from top-quark decays. The ratio of branching ratios 
\(R_{\tau/e}=\mathcal{B}(W\rightarrow\tau\nu_\tau)/\mathcal{B}(W\rightarrow e\nu_e)\) is determined by fitting the
transverse impact-parameter distributions of electrons, and is found to be compatible with one~\cite{TOPQ-2023-10}.
The CMS Collaboration tested lepton-flavour violation in \(Z\) boson decays, searching for events with the decays 
\(Z\rightarrow e\mu\), \(Z\rightarrow e\tau\) and \(Z\rightarrow\mu\tau\)~\cite{CMS-SMP-23-003}. Limits are set 
on the corresponding branching ratios at the level of \(10^{-5}\) to \(10^{-7}\). Samples enriched in events in which 
three or four top-quarks are produced are used to set limits on EFT coefficients, look for top-philic resonances and 
constrain the Yukawa coupling of the top quark~\cite{CMS-PAS-TOP-24-008}. The ATLAS Collaboration performed a search 
for same-sign top-quark production and interprets the results in terms of constraints on EFT 
coefficients~\cite{TOPQ-2021-14}, while the CMS Collaboration scrutinized variables sensitive to CP-odd EFT operators 
in samples of events which are dominated by \(t\bar{t}+Z\) production and \(tZq\) production~\cite{CMS:2025dpp}.
By combining two analyses measuring multi-lepton final states the sensitivity to the three EFT coefficients 
\(c_{tW}\), \(c_{\varphi Q}^3\) and \(c_{\phi t}\) is enhanced~\cite{CMS-PAS-TOP-24-004}. 
A comprehensive view at the limits on EFT coefficients is obtained by combining the CMS results of various analyses in 
the fields of Higgs boson, electroweak, top quark and quantum chromodynamics, with measurements of electroweak precision 
observables~\cite{CMS-SMP-24-003}.

\section{Pursuing subtlety and refinement}
\label{sec:subtlety}
The measurement of asymmetries allows to pursue subtle effects in top-quark production. The CMS Collaboration 
measured the lepton-charge asymmetry in \(t\bar{t}+W\) production~\cite{CMS-TOP-24-003}, obtaining a value compatible 
with the SM expectation. A new measurement determines the inclination asymmetry expressed with the angle \(\varphi\) 
between the planes of initial-state and final-state momenta~\cite{CMS:2025qmd}. 
The ATLAS Collaboration searched for the electroweak production of \(t\bar{t}+W+\,\)jet final states, 
corresponding to \(tW\) scattering processes which involve, for example, a \(Z\) boson or a Higgs boson 
as a mediator~\cite{TOPQ-2019-18}. While the observed data do not yield evidence 
for this process, limits on EFT contributions are set.  

Last year, the ATLAS and CMS Collaborations managed to establish the subtle quantum mechanical effect of quantum 
entanglement of \(t\bar{t}\) pairs produced near the kinematic threshold in the dileption 
channel~\cite{TOPQ-2021-24,CMS-TOP-23-001}. 
The possibility of entanglement of a set of particles is a fundamental feature of quantum mechanics 
which denotes the phenomenon that the quantum state of one particle cannot be described 
independently of the others. 
The new measurements demonstrate the occurrence of entanglement in a novel setting at high particle energy. 
The CMS Collaboration added an analysis in the lepton+jets channel 
that led to the observation of entanglement of \(t\bar{t}\) pairs with high invariant mass, 
namely \(m(t\bar{t})>800\,\)GeV~\cite{CMS-TOP-23-007}. 

\begin{figure}[!t]
\centering
\includegraphics[width=\textwidth]{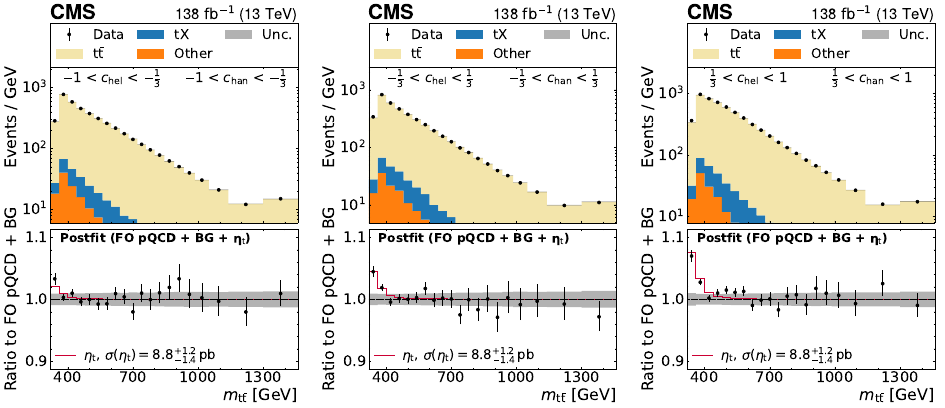}
\caption{Observed (points with statistical error bars) and predicted (stacked coloured histograms) 
\(m(t\bar{t})\) distribution in three out of nine bins of the two angular observables 
\(c_\mathrm{hel}\) and \(c_\mathrm{han}\) quantifying the degree of spin correlation 
between the top quark and the top antiquark. In the upper panels, the \(t\bar{t}\) 
histogram shows the fixed-order perturbative prediction in quantum chromodynamics 
after the fit to the data that includes the pseudoscalar \(\eta_t\) signal model 
(whose contribution is not drawn). The shown event rates are divided 
by the bin width. The lower panels display the ratio of the data to the prediction 
(without the pseudoscalar signal), with the \(\eta_t\) signal overlaid at its 
best fit cross-section (red line). The gray band indicates the postfit uncertainty.} 
\label{fig2}
\end{figure}

New analyses performed by the collaborations established quasi-bound-state effects leading to a cross-section enhancement 
in the \(m(t\bar{t})\) distribution near the kinematic threshold~\cite{ATLAS-CONF-2025-008,CMS-TOP-24-007}. This subtle effect could only be 
unravelled by exploiting the spin correlation between the top quark and top antiquark. 
Figure~\ref{fig2} shows the results obtained by the CMS Collaboration based on a pseudoscalar signal model 
\(\eta_t\).
Different signal models lead to slightly different signal strengths, but the cross-section 
enhancement remains.

\section{Conclusions}
Top-quark physics remains a vibrant field of basic research in particle physics. Recent new results 
comprise the first observation of scattering processes involving top quarks, 
namely \(t\bar{t}\gamma\gamma\) and \(tWZ\) production, and precision measurements of 
the top-quark mass and the \(t\bar{t}\) cross-section. Differential cross-section measurements 
of \(t\bar{t}\) production, \(tq\) production and \(t\bar{t}+W\) production extend the range to 
more extreme phase spaces. Direct and indirect searches use data sets with top quarks to 
scrutinise the SM, including searches for leptoquarks and for flavour-changing neutral currents, 
and tests of charged-lepton-flavour universality. In recent years, physicists pushed the frontier of 
challenging measurements further to establish subtle effects in top-quark production, such as 
the entanglement of \(t\bar{t}\) pairs in certain regions of phase space and 
quasi-bound-state effects leading to a cross-section enhancement in the kinematic threshold region.

\section*{Acknowledgements}
The author would like to thank the international advisory committee of the 
18\(^\mathrm{th}\) edition of the Workshop on Top-Quark Physics for inviting him to give the 
experimental summary talk. He extends his thanks to the local organising committee of 
Hanyang University in Seoul for being perfect hosts. The author acknowledges the financial 
support of the Federal Ministry of Research, Technology and Space of Germany (BMFTR).

\end{document}